\documentclass[vecphys]{svmult}

\usepackage{graphicx}        
\usepackage[bottom]{footmisc}
\usepackage{url}

\begin{document}

\title*{A disc in the heart of the Ant nebula}
\author{Foteini Lykou\inst{1}\and
Olivier Chesneau\inst{2}\and Eric Lagadec\inst{1}\and Albert Zijlstra\inst{1}}
\institute{University of Manchester, School of Physics \& Astronomy, P.O.Box 88, Manchester, M60 1QD, U.K.
\texttt{Foteini.Lykou@postgrad.manchester.ac.uk}
\and Observatoire de la C\^ote d'Azur, Dept. Gemini, Avenue N. Copernic, 06130, Grasse, France \texttt{olivier.chesneau@obs-azur.fr}}
%
%
\maketitle

\begin{abstract}
We present the discovery of a silicate disc at the centre of the planetary nebula Mz3 (the Ant). The nebula was observed with MIDI on the Very Large Telescope Interferometer (VLTI). The visibilities obtained at different orientations clearly indicate the presence of a dusty, nearly edge-on disc in the heart of the nebula. An amorphous silicate absorption feature is clearly seen in our mid-IR spectrum and visibility curves. We used radiative transfer Monte Carlo simulations to constrain the geometrical and physical parameters of the disc. We derive an inner radius of 9 AU ($\sim$6mas assuming D=1.4kpc). This disc is perpendicular to, but a factor of $10^{3}$ smaller than the optical bipolar outflow.

\end{abstract}

\section{Menzel 3}
\label{sec:1}
Complex phenomena perturb the mass-ejection in the late stages of stellar evolution: stellar magnetic fields, rotation or binarity are often invoked. Those mechanisms can lead to the creation of a circumstellar, dusty disc. The dust is created in the outer parts of the former stellar envelope. The Ant planetary nebula, or Mz3 is a bipolar nebula that has one of the most ``pinched'' waists. Its central object is suspected to be a binary but there are no stringent constraints. The temperature of the ionization source is about 30,000K and its distance is between $1-2$ kpc. Different expansion phases have shaped the current form of the nebula as seen in the Hubble Space Telescope images and long-slit spectroscopy \cite{gue,san}. Former studies \cite{smi1,smi2} have proposed the existence of a circumstellar disc, but due to its small size it could only be observed with high-angular resolution interferometry and only in the infrared.

\section{Infrared interferometry}
\label{sec:2}
The central objects of bipolar planetary nebulae are quite small compared to the PNe size. The ability of optical telescopes to detect these objects is very limited, while interferometers are quite suitable for the job. We have used the MIDI instrument of VLTI at ESO, which has a high-angular resolution $\sim0.01$ arcsec in the infrared. A combination of three different 8.2m Unit Telescopes, with the use of the mid-infrared recombiner MIDI operating in N band ($8-13.5 \mu m$), gave us six different baselines at different orientations across the Ant (Table~\ref{tab:1}). We have obtained the visibilities for Mz3 for each baseline and they revealed an edge-on disc of silicate dust around the central source. One should keep in mind that the higher the visibility, the smallest the source in the baseline direction. From all the above, we obtained constraints for the typical size and geometry of the N-band disc.

\begin{table}
\centering
\caption{VLTI baselines of Unit Telescopes 2, 3 and 4.}
\label{tab:1}       
\begin{tabular}{cc}
\hline\noalign{\smallskip}
B (m)   &  P.A. (deg) \\
\noalign{\smallskip}\hline\noalign{\smallskip}
46.3	& 1.5	\\
45.4	& 30.5	\\
31.4	& 73.8	\\
60.6	& 149.2	\\
52.0	& 77.2	\\
62.5	& 122.1	\\
\noalign{\smallskip}\hline
\end{tabular}
\end{table}

\begin{figure}
\centering
\includegraphics[height=8cm,width=\textwidth]{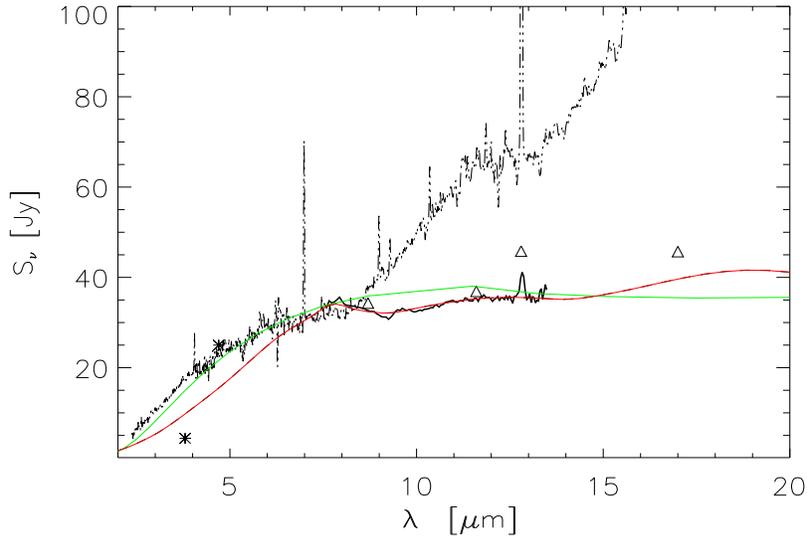}
\caption{SED of silicate (red) and graphite (green) models compared to the ISO (dotted) and MIDI spectra, Smith \& Gehrz (2005) points (triangles) and Chesneau et al. (2007) photometry (stars).}
\label{fig:1}       
\end{figure}

\section{The circumstellar disc}
\label{sec:3}
\begin{figure}
\centering
\includegraphics[height=8cm,width=\textwidth]{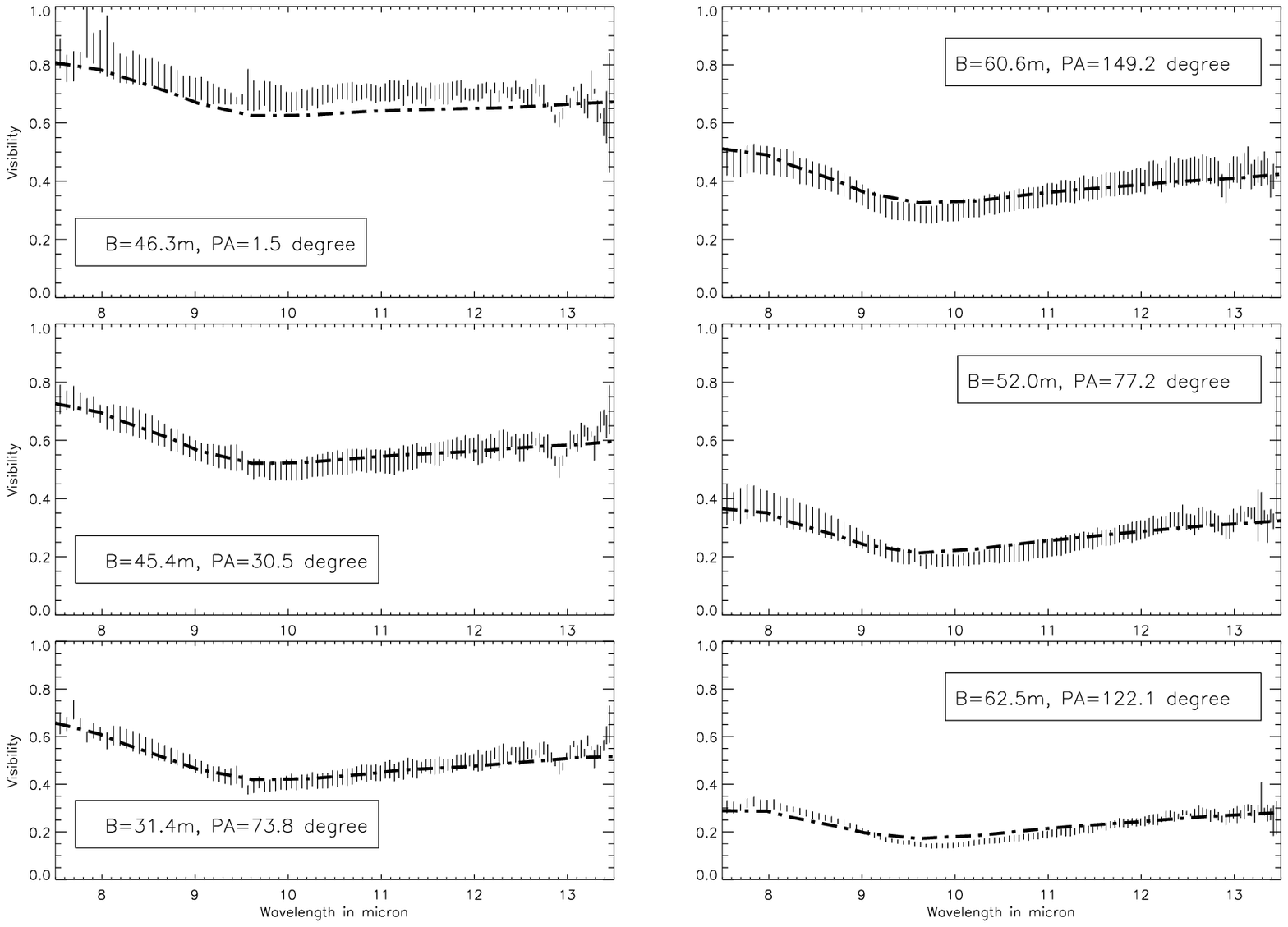}
\caption{Modeled visibility spectra (dash-dot) of amorphous silicates for every MIDI baseline.}
\label{fig:2}
\end{figure}

Radiative transfer simulations were made using the MC3D code \cite{wolf}. Figures~\ref{fig:1} and~\ref{fig:2}  present our comparisons between models and observations, showing that a model of a disc with a small aperture and high inclination matches almost perfectly the interferometric observations (mean $\chi^{2}=1.68$). The spectral energy distribution (SED) though, is not perfectly fitted; but this is just a beam effect, since the large beams used in these observations included emission from the bright lobes in all wavelengths. \par
\begin{figure}
\centering
\includegraphics[height=7.5cm]{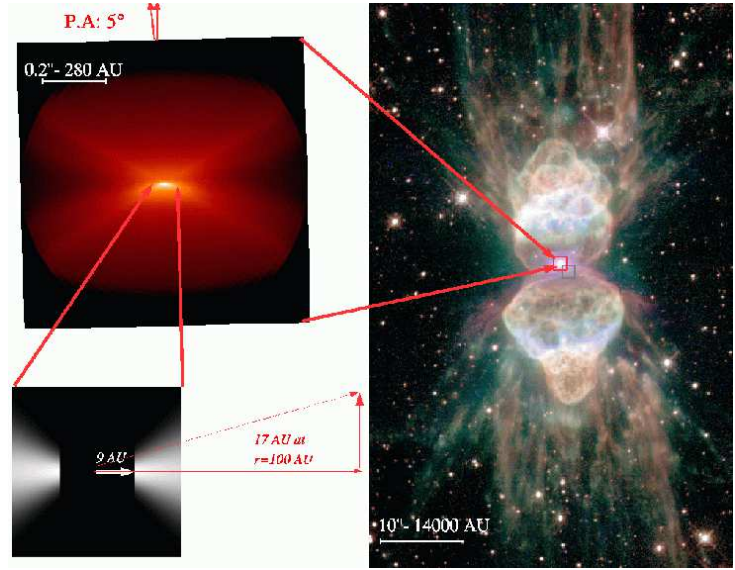}
\caption{Right: Menzel 3 imaged by the Hubble Space Telescope. Upper left: model of the 10$\mu$m flux distribution as deduced from MIDI observations. The lower part of the image (represents the southern lobe) is brighter, since this lobe is closer in our line-of-sight. Bottom left: modeled density cut of the inner part of the disk}
\label{fig:3}
\end{figure}

The disc parameters are the following: mass of the dust  $\sim10^{-5}M_{\odot}$ ; inclination of the disc $\sim74^{o}\pm3$ ; inner radius $9\pm1$AU respectively. We assumed that the distance to Mz3 is 1.4kpc, the effective temperature of the central source is 35,000K, the luminosity is $10^{4}L_{\odot}$ and that the grain size range is $0.05-1\mu m$. In addition, the disc (Figure~\ref{fig:3}) has a scale height of $17\pm2$ AU at 100AU from the star, with a moderate flaring (suggestive of a Keplerian structure?). It is worth mentioning that the mass of dust in the disc is 100 times smaller than that found in the lobes ($2.6\times10^{-3}M_{\odot}$~\cite{smi2}). This indicates that most of the ejected mass was channeled towards the lobes.

\section{Prospects}
\label{sec:4}
The current visibility measurements from MIDI are not able to provide a better definition for the size of the discs's inner rim and clarify whether the disc is circumstellar or circumbinary. New observations of Mz3 with the AMBER instrument of VLTI are expected to be completed in the near future and with these results the abovementioned questions will be answered. For more information please refer to Chesneau et al. 2007.



\end{document}